\documentclass[showpacs,amsmath,amssymb,aps,prd]{revtex4}

\usepackage{bm}% bold math

\begin{document}

\title{Differential rotation of nonlinear r-modes}

\author{Paulo M. S\'a}
\altaffiliation[Also at ]{Centro Multidisciplinar de
Astrof\'{\i}sica -- CENTRA/UAlg}
\email{pmsa@ualg.pt}
\affiliation{Departamento de F\'{\i}sica, F.C.T., Universidade do Algarve,
\\ Campus de Gambelas, 8005-139 Faro, Portugal}

\date{October 20, 2003}

\begin{abstract}
Differential rotation of r-modes is investigated within the nonlinear theory
up to second order in the mode amplitude in the case of a slowly-rotating,
Newtonian, barotropic, perfect-fluid star. We find a nonlinear extension of
the linear r-mode, which represents differential rotation that produces large
scale drifts of fluid elements along stellar latitudes. This solution includes
a piece induced by first-order quantities and another one which is a pure
second-order effect. Since the latter is stratified on cylinders, it cannot
cancel differential rotation induced by first-order quantities, which is not
stratified on cylinders. It is shown that, unlikely the situation in the
linearized theory, r-modes do not preserve vorticity of fluid elements at
second-order. It is also shown that the physical angular momentum and energy
of the perturbation are, in general, different from the corresponding
canonical quantities.
\end{abstract}


\maketitle

\section{\label{int}Introduction}

R-modes in Newtonian gravity were first studied more than twenty years ago
\cite{pp,pbr,sai}. In recent years, it was discovered numerically \cite{and},
and soon afterwards confirmed analytically \cite{fm}, that these modes,
analogous to Rossby waves in the Earth's atmosphere and oceans, are driven
unstable by gravitational radiation reaction in perfect fluid stars with
arbitrary small rotation.

The interest in the astrophysical implications of r-modes increased
dramatically when it was shown \cite{lom,aks} that in a newly born, hot,
rapidly-rotating neutron star the radiation reaction force dominates bulk and
shear viscosity for enough time to allow most of the star's angular momentum
to be radiated away as gravitational waves. As a result, the neutron star
spins down to just a small fraction of its initial angular velocity, thus
providing a possible explanation for the relatively small spin rates of young
pulsars in supernova remnants. For typical equations of state of a neutron
star it was estimated \cite{lom,aks} that the r-mode instability spins down a
young neutron star to a period of about 10--20 ms. This is comparable to the
inferred initial periods $P_0$ of the fastest pulsars associated with
supernova remnants, such as the Crab pulsar PSR B0531+21 ($P_0=19$ ms), or PSR
J0537-6910 ($P_0\ll 16$ ms for braking index $n=3$ \cite{mgzmw}). The
detectability of gravitational waves from such sources have been analyzed
within a simple phenomenological model for the evolution of the r-mode
instability, with the conclusion that gravitational waves emitted from an
young, rapidly rotating neutron star in the Virgo cluster could be detected by
enhanced versions of laser interferometer detectors \cite{olcsva}. However,
recent results on the nonlinear saturation of the r-mode energy indicate that
enhanced LIGO detectors could detect gravitational waves from these sources
only up to a distance of about 200 kpc \cite{afmstw}.

Another interesting astrophysical implication of r-modes is related to the
fact that gravitational-wave emission due to mode instability could balance
the spin-up torque due to accretion of neutron stars in low-mass X-ray
binaries \cite{bil,akst}, thus limiting the maximum angular velocity of these
stars to values consistent with observations (recent results indicate that the
fast end of the spin distribution could be around 600 Hz, well below the
theoretical maximum \cite{cmmgwkm}). If the viscosity of the neutron star is a
decreasing function of temperature, then the star undergoes a cyclical
evolution of spin-up due to accretion and spin-down due to gravitational
radiation emission \cite{l}. Assuming that the saturation amplitude of the
r-mode is of order unity, it was estimated \cite{l} that the time spent by the
star in the spin-down phase is about one year, just a small fraction of the
full duration of the cycle (several million years), thus making it unlikely
that any such source in our Galaxy is presently emitting gravitational waves.
However, recent results indicate that the saturation amplitude of the r-mode
could be much smaller than unity, implying that the duration of the spin-down
phase could be a much larger fraction of the cycle \cite{heyl}; in that case,
gravitational radiation from such sources in our galaxy could be detected by
enhanced LIGO detectors \cite{afmstw}.

While initial research on r-modes was carried out within the linearized
theory, a deeper understanding of these modes and its astrophysical
implications requires considering the nonlinear theory.

An important nonlinear effect, that has been investigated recently by several
authors, is differential rotation induced by r-modes. This differential
rotation could wind up a magnetic field in an old, accreting neutron star in
an X-ray binary, leading to a gamma-ray burst \cite{sp}. Differential rotation
could also interact with the magnetic field of a newly born, rapidly rotating
neutron star, limiting the growth of the r-mode instability or, for strong
magnetic fields, even preventing it from developing \cite{rls,rlms1,rlms2}.

An analytical expression for differential rotation induced by r-modes was
first derived using the linearized fluid equations by expanding the velocity
of a fluid element located at a certain point in powers of the mode amplitude,
averaging over a gyration, and retaining only the lowest-order nonvanishing
term \cite{rls}. Since this procedure is not equivalent to solving the
nonlinear fluid equations, the analytical expression thus obtained is just an
approximate one. Differential rotation was also reported in a useful toy model
of a thin spherical shell of a rotating incompressible fluid \cite{lu}.
However, in this case, differential rotation is of a different nature than the
one referred to above; it is driven by radiation reaction. The existence of
differential rotation related to r-modes was confirmed by numerical
simulations \cite{sf,ltv}.

In this paper, differential rotation induced by r-modes is investigated within
the nonlinear theory up to second order in the mode amplitude $\alpha$ in the
case of a Newtonian, barotropic, perfect-fluid star rotating with constant
angular velocity $\Omega$. Second-order quantities are obtained by expanding
simultaneously in powers of $\alpha$ and $\Omega$. In order to consistently
neglect higher-order terms arising from the expansion in $\Omega$, it is
assumed that $\alpha \gg (\Omega/\Omega_K)^2$, where $\Omega_K$ is the angular
velocity at which the star starts shedding mass through the equator. The
investigation of r-modes up to second order in the angular velocity $\Omega$,
neglecting higher-order terms arising from the expansion in $\alpha$, was
carried out in Ref.~\cite{aks} (using Saio's formalism \cite{sai}) and in
Ref.~\cite{lmo} (using the two-potential formalism \cite{il}). In order to
neglect the deformation of the star due to the centrifugal force, our analysis
is carried out in a slow-rotation approximation, $\Omega \ll \Omega_K$.

In section \ref{lin} we review some important results obtained in the linear
theory of r-modes. In section \ref{sec}, a nonlinear extension of the linear
r-mode perturbation is derived up to second order in the mode amplitude
$\alpha$. This solution describes differential rotation that produces large
scale drifts of fluid elements along stellar latitudes. The Lagrangian theory
of nonrelativistic fluids \cite{fs} is used in this section to apply the
boundary condition at the surface of the star and to show that, at
second-order, r-modes do not preserve vorticity of fluid elements. The
physical angular momentum and energy of the second-order r-mode solution are
analyzed in section \ref{ang}. In particular, it is shown that, in general,
physical and canonical quantities are different. Finally, section \ref{dis} is
devoted to discussion and conclusions.

\section{\label{lin}R-modes in the linearized theory}

The linearized fluid equations for an uniformly rotating, Newtowian,
barotropic, perfect-fluid star in an inertial frame are
\begin{eqnarray}
&
 \partial_t \delta^{(1)} v_i
+\delta^{(1)} v^k \nabla_k v_i
+v^k \nabla_k \delta^{(1)} v_i
=-\nabla_i \left( \frac{\delta^{(1)} p}{\rho} + \delta^{(1)} \Phi \right),
&
\label{eulerp1}
\\
&
 \partial_t \delta^{(1)}\rho
+v^i \nabla_i \delta^{(1)}\rho
+\nabla_i (\rho \delta^{(1)} v^i)=0,
&
\label{contp1}
\\
&
\nabla^i \nabla_i \delta^{(1)} \Phi=4\pi G\delta^{(1)}\rho,
&
\label{poisp1}
\end{eqnarray}
where $v^i=\Omega \delta_{\phi}^i$ and $\rho$ are, respectively, the fluid
velocity and the mass density of the unperturbed star, $\rho$ is related to
the pressure $p$ by an equation of state of the form $p=p(\rho)$, and the
quantities $\delta^{(1)} v^i$, $\delta^{(1)} p$, $\delta^{(1)}\rho$, and
$\delta^{(1)}\Phi$ are, respectively, the first-order Eulerian change in
velocity, pressure, density, and gravitational potential.

At lowest order in $\Omega$ the linearized Euler equation (\ref{eulerp1})
yields, in spherical coordinates $(r,\theta,\phi)$, the r-mode solution
\begin{subequations}
\label{vf}
\begin{eqnarray}
\delta^{(1)} v^r &=& 0,
    \label{vf1} \\
\delta^{(1)} v^{\theta} &=& \alpha \Omega C_{l} l (r/R)^{l-1}
    \sin^{l-1}\theta \sin (l\phi+\omega t),
    \label{vf2} \\
\delta^{(1)} v^{\phi} &=& \alpha \Omega  C_{l} l (r/R)^{l-1}
    \sin^{l-2}\theta \cos\theta \cos (l\phi+\omega t)
    \label{vf3},
\end{eqnarray}
\end{subequations}
with $\delta^{(1)} U \equiv \delta^{(1)} p/\rho +\delta^{(1)} \Phi$
given by
\begin{equation}
\delta^{(1)} U=2 \alpha \Omega^2  \frac{C_l l}{l+1}
R^2 \left( \frac{r}{R} \right)^{l+1}
\sin^l\theta\cos\theta \cos(l\phi+\omega t),
\label{mp1}
\end{equation}
where $\omega=-\Omega l+2\Omega/(l+1)$ is the mode angular frequency in the
inertial frame, $R$ is the radius of the unperturbed star and
$C_l=(2l-1)!!\sqrt{(2l+1)/[2\pi(2l)!l(l+1)]}$. We will only consider modes for
which $l\geq2$.

The above solution satisfies the linearized continuity equation
(\ref{contp1}), which at lowest order in $\Omega$ is simply $\nabla_i (\rho
\delta^{(1)} v^i )=0$.

The linearized Euler and continuity equations do not contain information on
how to split $\delta^{(1)}U$ into the first-order changes in pressure
$\delta^{(1)}p$ and gravitational potential $\delta^{(1)}\Phi$. This has to be
done by using the perturbed Poisson equation (\ref{poisp1}) and the boundary
conditions for the gravitational potential at the surface of the star and at
infinity. Assuming for $\delta^{(1)} \Phi$ the same angular dependence as
$\delta^{(1)}U$, its (dimensionless) radial part $f(r)$ must be a solution of
the equation \cite{lmo}
\begin{equation}
\frac{d^2 f(r)}{dr^2}+\frac{2}{r} \frac{df(r)}{dr}
+\left( 4\pi G\rho \frac{d\rho}{dp}
-\frac{(l+1)(l+2)}{r^2}\right)f(r)=
4\pi G \rho\frac{d\rho}{dp} \left( \frac{r}{R} \right)^{l+1},
\end{equation}
with the conditions that $\delta^{(1)} \Phi$ be continuous at the surface of
the star, have a continuous first derivative there and tend to zero as
$r\rightarrow\infty$.

The r-mode solution given by eqs.~(\ref{vf}) and (\ref{mp1}) also satisfies
the zero-boundary condition for the pressure $p$ at the surface of the star,
i.e., the first-order Lagrangian change in pressure \footnote{
        Throughout this article, we use the definitions and notation of
        Ref.~\cite{fs}.
        In particular, the first-order Lagrangian change in a quantity $q$
        is defined as
        $\Delta_{\xi}^{(1)} q=\delta^{(1)} q+ \pounds_{\xi^{(1)}} q$,
        where $\xi^{(1)}$ is the first-order Lagrangian displacement vector
        which connects fluid elements in the equilibrium with
        corresponding ones in the perturbed configuration;
        the action of the Lie derivative $\pounds_{\xi^{(1)}}$ on a tensor
        ${T^{a_1\dots a_k}}_{b_1\dots b_l}$ is given by
        $\pounds_{\xi^{(1)}} {T^{a_1\dots a_k}}_{b_1\dots b_l}=
        \xi^{(1)c}\nabla_c  {T^{a_1\dots a_k}}_{b_1\dots b_l}
        -\sum_{i=1}^k {T^{a_1\dots c\dots a_k}}_{b_1\dots b_l}
        \nabla_c \xi^{(1)a_i}
        +\sum_{j=1}^l {T^{a_1\dots a_k}}_{b_1\dots c\dots b_l}
        \nabla_{b_j} \xi^{(1)c}$.
        Note that the above definition of Lagrangian change agrees
        with the one commonly used in some literature of stellar pulsation,
        $\Delta_{\xi}^{(1)} q=\delta_{\xi}^{(1)} q+ \xi^{(1)a}\nabla_a q$,
        only for scalars.
        }
vanishes at the surface of the star, $\Delta_{\xi}^{(1)} p=0$.

The Lagrangian change in pressure is, at first order, given by
\begin{equation}
\Delta_{\xi}^{(1)} p= -\gamma p \nabla_i \xi^{(1)i} \label{bcp1},
\end{equation}
where $\xi^{(1)i}$ are the contravariant components of the first-order
Lagrangian displacement vector and
$\gamma=(\rho/p)(\partial p/\partial\rho)$.
These components are the solution of the equations \cite{fs}
\begin{eqnarray}
\delta^{(1)} v^i&=&\partial_{t} \xi^{(1)i}+v^k\nabla_k\xi^{(1)i}
           -\xi^{(1)m}\nabla_m v^i
          =\left( \partial_{t}+\Omega\partial_{\phi} \right)\xi^{(1)i},
\label{xivp1}
\\
\delta^{(1)} \rho&=&-\nabla_i(\rho \xi^{(1)i}),
\label{xirhop1}
\end{eqnarray}
where $\partial_{t}+\Omega\partial_{\phi}$ is the time derivative in the
rotating frame, $\partial_{t'}$. At lowest order in $\Omega$ these equations
yield for $\xi^{(1)i}$ the following solution:
\begin{subequations}
\label{ld}
\begin{eqnarray}
\xi^{(1)r}&=&0,  \label{ld1}
\\
\xi^{(1)\theta}&=&
- \frac12 \alpha C_l l(l+1) \left( \frac{r}{R} \right)^{l-1}
\sin^{l-1} \theta
\cos (l\phi+\omega t),  \label{ld2}
\\
\xi^{(1)\phi}&=&
\frac12 \alpha C_l l(l+1) \left( \frac{r}{R} \right)^{l-1}
\sin^{l-2}\theta \cos\theta
\sin(l\phi+\omega t).  \label{ld3}
\end{eqnarray}
\end{subequations}
Using the above $\xi^{(1)i}$ it is now straightforward to verify that the
Lagrangian change in pressure $\Delta_{\xi}^{(1)} p$ vanishes everywhere in
the star, including the surface.

The Lagrangian displacement $\xi^{(1)i}$, given by eqs.~(\ref{ld}), is
canonical \cite{fs}; it satisfies the conditions $\epsilon^{ijk}\nabla_j
\Delta_{\xi}^{(1)} v_k=0, \label{cc}$ where the first-order Lagrangian change
in velocity is given by
\begin{equation}
\Delta_{\xi}^{(1)} v_i=\partial_{t} \xi^{(1)}_i
+v^k \left( \nabla_i\xi^{(1)}_k + \nabla_k \xi^{(1)}_i \right).
\end{equation}
For the canonical Lagrangian displacement (\ref{ld}) the canonical energy
\begin{eqnarray}
E_c&=&\frac12\int\left[
\rho \partial_{t} \xi^{(1)i} \partial_{t} \xi^{(1)}_i
-\rho v^j \nabla_j \xi^{(1)i} v^k \nabla_k \xi^{(1)}_i
+\gamma p(\nabla_i\xi^{(1)i})^2
+2\xi^{(1)i} \nabla_i p \nabla_j \xi^{(1)j} \right.
\nonumber
\\
& & {} + \left. \xi^{(1)i} \xi^{(1)j} (\nabla_i \nabla_j p
+ \rho \nabla_i \nabla_j \Phi)
\right] dV +{\cal O}(\Omega^4),
\label{ce}
\end{eqnarray}
is negative for any $l\geq 2$ and for arbitrarily slow rotation \cite{fm},
implying r-modes instability to gravitational radiation \cite{fs2}. Let us
emphasize here that the use of this stability criterion does not require
knowledge of second-order Lagrangian displacements, since the canonical energy
is quadratic in first-order quantities.

Using the relations
$\Delta_{\xi}^{(1)}\epsilon^{ijk}=-\epsilon^{ijk}\nabla_m\xi^{(1)m}=0$ and
$\Delta_{\xi}^{(1)}\nabla_j v_k = \nabla_j\Delta_{\xi}^{(1)} v_k -
v^m\nabla_j\nabla_k\xi_m^{(1)}$, it immediately follows that the canonical
condition $\epsilon^{ijk}\nabla_j\Delta_{\xi}^{(1)}v_k=0$ is equivalent to the
statement that the first-order Lagrangian change in vorticity is zero,
$\Delta_{\xi}^{(1)}(\epsilon^{ijk}\nabla_j v_k)=0$. Thus, canonical
displacements are precisely those that preserve vorticity. If the vorticity of
a r-mode perturbation is initially small, then as the perturbation grows under
gravitational radiation reaction the vorticity will not grow.

\section{\label{sec}The second-order r-mode solution}

At second order, the perturbed Euler, continuity, and Poisson equations in an
inertial frame are given, respectively, by
\begin{eqnarray}
&
 \partial_t \delta^{(2)} v_i
+\delta^{(2)} v^k \nabla_k v_i
+v^k \nabla_k \delta^{(2)} v_i
+\delta^{(1)}v^k \nabla_k \delta^{(1)} v_i
=-\nabla_i \delta^{(2)} U
+\frac{\delta^{(1)}\rho}{\rho} \nabla_i
  \left( \frac{\delta^{(1)} p}{\rho} \right),
&
\label{eulerp2}
\\
&
 \partial_t \delta^{(2)}\rho
+v^i \nabla_i \delta^{(2)}\rho
+\nabla_i (\rho \delta^{(2)} v^i)
+\nabla_i (\delta^{(1)}\rho  \delta^{(1)} v^i) = 0,
&
\label{contp2}
\\
&
\nabla^i \nabla_i \delta^{(2)} \Phi=4\pi G\delta^{(2)}\rho,
&
\label{poisp2}
\end{eqnarray}
where $\delta^{(2)} U\equiv \delta^{(2)} p/\rho + \delta^{(2)} \Phi$
and $\delta^{(2)} q$ denotes the second-order Eulerian change
in a quantity $q$.

Let us now assume that $\alpha \gg (\Omega/\Omega_K)^2$. Then, terms in
$\delta^{(2)} v^i$ proportional to $\alpha\Omega^3$ (arising in an expansion
in powers of the angular velocity of the star) can be neglected when compared
with terms proportional to $\alpha^2\Omega$ (arising in an expansion in powers
of the mode amplitude). For the same reason, we neglect in $\delta^{(2)} U$
and $\delta^{(2)} \rho$ terms proportional to $\alpha\Omega^4$. Since the
second term on the right-hand side of eq.~(\ref{eulerp2}) is of order
$\alpha^2\Omega^4$, equation (\ref{eulerp2}) reduces, at lowest order in
$\Omega$, to
\begin{equation}
 \partial_t \delta^{(2)} v_i
+\delta^{(2)} v^k \nabla_k v_i
+v^k \nabla_k \delta^{(2)} v_i
+\nabla_i \delta^{(2)} U
=-\delta^{(1)}v^k \nabla_k \delta^{(1)} v_i,
\label{eulerp2a}
\end{equation}
where $\delta^{(1)} v^i$ is of order $\alpha\Omega$, $\delta^{(2)} v^i$ is of
order $\alpha^2\Omega$ and $\delta^{(2)} U$ is of order $\alpha^2\Omega^2$. In
this equation second-order quantities depend on the first-order ones only
through the term quadratic on $\delta^{(1)} v^i$. Using eqs.~(\ref{vf}) to
compute this term, the above equation reads, in spherical coordinates
$(r,\theta,\phi)$,
\begin{subequations}
\label{eulerp2b}
\begin{eqnarray}
& &(\partial_t+\Omega\partial_\phi) \delta^{(2)} v^r
  -2\Omega r\sin^2\theta \delta^{(2)} v^{\phi}
  +\partial_r \delta^{(2)} U =
\nonumber \\
& &\quad -\frac12\alpha^2\Omega ^2 C_l^2 l^2 R
              \left( \frac{r}{R} \right)^{2l-1}
   \sin^{2l-2}\theta
   \left\{ \sin^2\theta-2+\sin^2\theta\cos[2(l\phi+\omega t)]
   \right\},
  \label{eulerp2b1}
\\
& &(\partial_t+\Omega\partial_\phi)  \delta^{(2)} v^{\theta}
  -2\Omega\sin\theta\cos\theta \delta^{(2)} v^{\phi}
  +\frac{1}{r^2} \partial_{\theta} \delta^{(2)} U =
\nonumber \\
& &\quad  -\frac12 \alpha^2 \Omega ^2 C_l^2 l^2
               \left( \frac{r}{R} \right)^{2l-2}
   \sin^{2l-3}\theta\cos\theta
   \left\{
   \sin^2\theta+2l-2+\sin^2\theta \cos[2(l\phi+\omega t)]
   \right\},
 \label{eulerp2b2}
\\
& &(\partial_t+\Omega\partial_\phi) \delta^{(2)} v^{\phi}
  +2\Omega \frac{\delta^{(2)} v^{r}}{r}
  +2\Omega \frac{\cos\theta}{\sin\theta} \delta^{(2)} v^{\theta}
  +\frac{1}{r^2\sin^2\theta} \partial_{\phi} \delta^{(2)} U =
\nonumber \\
& &\quad + \frac12 \alpha^2 \Omega ^2 C_l^2 l^2
                \left( \frac{r}{R} \right)^{2l-2}
   \sin^{2l-2}\theta \sin[2(l\phi+\omega t)].
 \label{eulerp2b3}
\end{eqnarray}
\end{subequations}
The right-hand side of the previous equations contains a piece that does not
dependent on $t$ and $\phi$ and a double ``frequency" $2(l\phi+\omega t)$
piece. The former induces an axisymmetric time-independent second-order
solution corresponding to differential rotation, while the latter induces a
second-order solution corresponding to an oscillating response at a
``frequency" twice that of the r-mode. In this article we will be concerned
exclusively with the solution corresponding to differential rotation of
r-modes.

The axisymmetric time-independent right-hand side of system~(\ref{eulerp2b})
induces the following second order solution:
\begin{subequations}
\label{difrot}
\begin{eqnarray}
\delta^{(2)} v^r&=& 0,  \label{difrot1}
\\
\delta^{(2)} v^{\theta}&=&0,  \label{difrot2}
\\
\delta^{(2)} v^{\phi} &=&
  \frac12 \alpha^2 \Omega C_l^2 l^2(l^2-1)
  \left(\frac{r}{R} \right)^{2l-2}
  \sin^{2l-4}\theta,
  \label{difrot3}
\end{eqnarray}
\end{subequations}
and
\begin{equation}
\delta^{(2)} U=-\frac14 \alpha^2\Omega^2 C_l^2 l R^2
          \left(\frac{r}{R} \right)^{2l} \sin^{2l-2}\theta
          \left[ \sin^2\theta -2l^2\right].
\label{partU}
\end{equation}
Indeed, assuming axisymmetry and time-independence system~(\ref{eulerp2b})
decouples into two independent systems, one determining $\delta^{(2)}
v^{\phi}$ and $\delta^{(2)} U$, the other relating $\delta^{(2)} v^{r}$ and
$\delta^{(2)} v^{\theta}$. Eliminating $\delta^{(2)} U$ from
eqs.~(\ref{eulerp2b1}) and (\ref{eulerp2b2}), one obtains an equation that
determines $\delta^{(2)} v^{\phi}$, namely
\begin{equation}
  \Omega \sin\theta \cos\theta \partial_r
    \left( r^2 \delta^{(2)} v^{\phi} \right)
 -\Omega r \partial_\theta
    \left( \sin^2\theta \delta^{(2)} v^{\phi} \right)
 =\alpha^2\Omega^2 C_l^2 l^2 (l^2-1)
                R \left( \frac{r}{R} \right)^{2l-1}
                \sin^{2l-3}\theta \cos\theta,
\label{u2a}
\end{equation}
yielding (\ref{difrot3}). Now, inserting the above solution for $\delta^{(2)}
v^{\phi}$ in eqs.~(\ref{eulerp2b1}) and (\ref{eulerp2b2}) one obtains
(\ref{partU}). Finally, through eq.~(\ref{eulerp2b3}), $\delta^{(2)} v^{r}=0$
implies $\delta^{(2)} v^{\theta}=0$.

The above second-order solution, induced by first-order quantities, describes
a drift of fluid elements along stellar latitudes. Note, however, that the
homogeneous part of system~(\ref{eulerp2b}), involving only second-order
quantities, also admits a solution describing a drift along stellar latitudes,
\begin{subequations}
\label{stra}
\begin{eqnarray}
&&\delta^{(2)} \hat{v}^{r}=0,
\label{stra1}
\\
&&\delta^{(2)} \hat{v}^{\theta}=0,
\label{stra2}
\\
&&\delta^{(2)} \hat{v}^{\phi} = \alpha^2 \Omega A
r^{N-1} \sin^{N-1}\theta,
\label{stra3}
\end{eqnarray}
\end{subequations}
with
\begin{equation}
\delta^{(2)} \hat{U}=
\frac{2\alpha^2 \Omega^2 A}{N+1} r^{N+1} \sin^{N+1}\theta,
\label{stra4}
\end{equation}
where $A$ and $N$ are some constants determined by initial data. Thus, the
full second-order solution describing differential rotation has a piece
induced by first-order quantities and another one which is a pure second-order
effect. Since the differential rotation given by eq.~(\ref{stra3}) is
stratified on cylinders, it cannot cancel differential rotation induced by
first-order quantities, which is \textit{not} stratified on cylinders. Thus,
differential rotation is an unavoidable feature of nonlinear r-modes.

The solution given by eqs.~(\ref{difrot}), (\ref{partU}),
(\ref{stra}) and (\ref{stra4})
is the nonlinear extension of the linear r-mode perturbation
we were looking for.

The above solution also satisfies the perturbed continuity equation, which at
lowest order in $\Omega$ is simply $\nabla_i (\rho \delta^{(2)} v^i)=0$.
Again, as in the linearized theory, the splitting of $\delta^{(2)} U$ into
$\delta^{(2)} p$ and $\delta^{(2)}\Phi$ is done by using the perturbed Poisson
equation (\ref{poisp2}) with appropriate boundary conditions for the
gravitational potential $\delta^{(2)}\Phi$.

The second-order r-mode solution should also satisfy the zero-boundary
condition for the pressure $p$ at the surface of the star, i.e., the
second-order Lagrangian change in pressure \footnote{
        In this section, the second-order Lagrangian change in a
        quantity $q$ is derived using the formalism developed in Appendix B
        of Ref.~\cite{fs}, taking into account that it contains,
        besides terms quadratic in the first-order Lagrangian displacement
        $\xi^{(1)i}$, terms linear in the second-order Lagrangian
        displacement $\xi^{(2)i}$ \cite{sc}.
        For scalars $f$ and contravariant vector fields $v^i$ the relation
        between the second-order Eulerian and Lagrangian changes is given,
        respectively, by
        $
        \Delta^{(2)}_{\xi}f=\delta^{(2)}f+\xi^{(2)a}\nabla_a f
        +1/2 \xi^{(1)a} \xi^{(1)b} \nabla_a\nabla_b f
        +\xi^{(1)a}\nabla_a \delta^{(1)}f
        $
        and
        $
        \Delta^{(2)}_{\xi}v^i=\delta^{(2)}v^i
        +\xi^{(2)a}\nabla_a v^i-v^a \nabla_a \xi^{(2)i}
        -\nabla_a \xi^{(1)i} (\xi^{(1)b} \nabla_b v^a-v^b \nabla_b
        \xi^{(1)a})
        +\xi^{(1)a} \nabla_a \delta^{(1)}v^i
        -\delta^{(1)}v^a \nabla_a \xi^{(1)i}.
        $
        }
should vanish at the surface of the star, $\Delta_{\xi}^{(2)} p=0$.

The Lagrangian change in pressure is given by
\begin{equation}
\Delta_\xi p = \gamma p \frac{\Delta_\xi\rho}{\rho}
   +\frac{\gamma (\gamma-1)}{2}p
    \left( \frac{\Delta_\xi\rho}{\rho} \right)^2+\dots,
\end{equation}
where $\gamma=(\rho/p)(\partial p/\partial\rho)$. The Lagrangian change in
density $\Delta_\xi \rho$ can be expressed in terms of the displacement vector
using conservation of mass \cite{fs}:
\begin{equation}
 \frac{\Delta_\xi\rho}{\rho}=-\nabla_k \xi^{(1)k}
     +\frac12 \left( \nabla_k \xi^{(1)k} \nabla_m \xi^{(1)m}
                    +\nabla_k \xi^{(1)m} \nabla_m \xi^{(1)k}\right)
     -\nabla_k \xi^{(2)k}+{\cal O}(\xi^3),
\label{Deltarho}
\end{equation}
where $\xi^{(2)i}$ are the contravariant components of the second-order
Lagrangian displacement vector.

Using eq.~(\ref{Deltarho}) the second-order Lagrangian change in pressure
$\Delta_\xi^{(2)} p$ is given by
\begin{equation}
\Delta_\xi^{(2)} p =
\gamma p \left( \frac12 \nabla_k \xi^{(1)m} \nabla_m \xi^{(1)k}
-\nabla_k \xi^{(2)k} \right)
+\frac{\gamma^2}{2}p \nabla_k \xi^{(1)k} \nabla_m \xi^{(1)m}.
\end{equation}
Since $\nabla_k \xi^{(1)k}=0$,
the boundary condition $\Delta_\xi^{(2)} p=0$
is satisfied if
\begin{equation}
\nabla_k \xi^{(2)k}=\frac12 \nabla_k \xi^{(1)m} \nabla_m \xi^{(1)k},
\end{equation}
or, using eqs.~(\ref{ld}), if
\begin{equation}
\nabla_k \xi^{(2)k}=
   \frac14 \alpha^2 C_l^2 l^2 (l+1)^2
    \left( \frac{r}{R} \right)^{2l-2}
        \left[ (l^2-l+1)\cos^2\theta-l \right]
     \sin^{2l-4}\theta.
\label{qaz1}
\end{equation}

The contravariant components of the second-order Lagrangian displacement
vector $\xi^{(2)i}$ are determined by the equations
\begin{eqnarray}
\delta^{(2)} v^i  &=&
(\partial_{t}+\Omega\partial_{\phi})\xi^{(2)i}
-\xi^{(1)k} \nabla_k\delta^{(1)} v^i,
\label{vi2}
\\
\delta^{(2)}\rho &=&
-\rho \left( \nabla_k \xi^{(2)k}
           -\frac12 \nabla_k \xi^{(1)m} \nabla_m \xi^{(1)k} \right)
-\left( \xi^{(2)k}\nabla_k\rho
           +\frac12\xi^{(1)k}\xi^{(1)m}\nabla_k\nabla_m\rho  \right),
\label{rho2}
\end{eqnarray}
where $\partial_{t}+\Omega\partial_{\phi}$ is the time derivative in the
rotating frame, $\partial_{t'}$. In the above equations, terms quadratic in
the first-order quantities $\xi^{(1)i}$ and $\delta^{(1)} v^i$ give rise to
double ``frequency" terms; since in this article we are concerned only with
differential rotation, these double frequency terms will not be considered.
Thus, at lowest order in $\Omega$, eq.~(\ref{vi2}) yields the solution:
\begin{subequations}
\label{edc}
\begin{eqnarray}
\xi^{(2)r} &=& C_1(r,\theta),
\label{edc1}
\\
\xi^{(2)\theta} &=& C_2(r,\theta),
\label{edc2}
\\
\xi^{(2)\phi} &=& D(r,\theta) t' + C_3(r,\theta), \label{edc3}
\end{eqnarray}
\end{subequations}
where
\begin{eqnarray}
D(r,\theta)&=&\frac14 \alpha^2 \Omega C_l^2 l^2 (l+1)(2l-1)
\left( \frac{r}{R} \right)^{2l-2} \sin^{2l-2}\theta
+ \alpha^2 \Omega A r^{N-1} \sin^{N-1}\theta,
\end{eqnarray}
and $C_i$ are some arbitrary functions of $r$ and $\theta$.

The second-order Lagrangian displacement vector $\xi^{(2)i}$, given by
eqs.~(\ref{edc}), is also a solution to eq.~(\ref{rho2}) at lowest order in
$\Omega$, if the functions $C_i$ are chosen to be
\begin{subequations}
\label{c}
\begin{eqnarray}
C_1(r,\theta) &=& \frac{1}{16} \alpha^2 C_l^2 l^2 (l+1)^2
R \left( \frac{r}{R} \right)^{2l-1} \sin^{2l-2}\theta
\left( \sin^2\theta-2 \right),
\\
\label{c1}
C_2(r,\theta) &=& \frac{1}{16} \alpha^2 C_l^2 l^2 (l+1)^2
\left( \frac{r}{R} \right)^{2l-2} \sin^{2l-3}\theta \cos\theta
\left( \sin^2\theta+2l-2 \right),
\label{c2}
\\
C_3(r,\theta) &=& 0.
\label{c3}
\end{eqnarray}
\end{subequations}
It is now straightforward to verify that $\xi^{(2)i}$, given by
eqs.~(\ref{edc})--(\ref{c}), satisfies the condition (\ref{qaz1}) everywhere
in the star, including the surface. Thus, at second-order the solution
satisfies the boundary condition $\Delta_\xi^{(2)} p=0$ for any value of the
constants $A$ and $N$.

At second order, the Lagrangian change in vorticity is given by
\begin{equation}
q^i \equiv\Delta_{\xi}^{(2)}(\epsilon^{ijk}\nabla_j v_k)=
\Delta_{\xi}^{(2)}\epsilon^{ijk} \nabla_j v_k+
\epsilon^{ijk} \Delta_{\xi}^{(2)} \nabla_j v_k+
\Delta_{\xi}^{(1)}\epsilon^{ijk} \Delta_{\xi}^{(1)} \nabla_j v_k,
\end{equation}
where
\begin{eqnarray}
\Delta_{\xi}^{(2)}\epsilon^{ijk}&=&
-\epsilon^{ijk} \nabla_m \xi^{(2)m}
-\xi^{(1)n}\nabla_n\left( \epsilon^{ijk}\nabla_m \xi^{(1)m} \right)
+\epsilon^{ijk} \nabla_m \left( \xi^{(1)n} \nabla_n \xi^{(1)m} \right)
\nonumber
\\
&& + \epsilon^{imn} \nabla_m \xi^{(1)j} \nabla_n \xi^{(1)k}+
\epsilon^{mjn} \nabla_m \xi^{(1)i} \nabla_n \xi^{(1)k}+
\epsilon^{mnk} \nabla_m \xi^{(1)i} \nabla_n \xi^{(1)j}
\label{mnb1}
\end{eqnarray}
and
\begin{equation}
\Delta_{\xi}^{(2)} \nabla_j v_k= \nabla_j \Delta_{\xi}^{(2)} v_k -v^m \nabla_j
\nabla_k \xi_m^{(2)} -\left(\partial_{t} \xi^{(1)m} + v^n \nabla_n \xi^{(1)m}
\right) \nabla_j \nabla_k \xi_m^{(1)}. \label{mnb2}
\end{equation}
Using $\xi^{(1)i}$ and $\xi^{(2)i}$ given, respectively, by
eqs.~(\ref{ld}) and eqs.~(\ref{edc})--(\ref{c}),
and taking into account that the second-order Lagrangian change in velocity
is given by \cite{fs}
\begin{equation}
\Delta^{(2)}_{\xi}v_i
=\partial_{t} \xi^{(1)k} \nabla_i \xi^{(1)}_k
+v^k \nabla_k \xi^{(1)m}\nabla_i \xi^{(1)}_m
+\partial_{t} \xi^{(2)}_i+v^k
    \left( \nabla_i \xi^{(2)}_k+\nabla_k \xi^{(2)}_i \right),
\label{lc2v}
\end{equation}
one obtains for $q^i$ the following expressions:
\begin{subequations}
\label{vrt}
\begin{eqnarray}
q^r &=& \frac14\alpha^2\Omega C_l^2 l^2(l+1)
        \left( \frac{r}{R} \right)^{2l-2}
      \sin^{2l-4}\theta\cos\theta
      [l(l+1)\sin^2\theta+4(l-1)^2]
\nonumber
\\
& & + \alpha^2 \Omega A (N+1) r^{N-1} \sin^{N-1}\theta\cos\theta,
\label{vrt1}
\\
q^{\theta} &=& -\frac14\alpha^2\Omega C_l^2 l^3 (l+1) R^{-1}
\left( \frac{r}{R} \right)^{2l-3}\sin^{2l-3}\theta
[(l+1)\sin^2\theta+4(l-1)]
\nonumber
\\
& & - \alpha^2 \Omega A (N+1) r^{N-2} \sin^N \theta,
\label{vrt2}
\\
q^{\phi} &=& 0.
\label{vrt3}
\end{eqnarray}
\end{subequations}

As it can be seen from eqs.~(\ref{vrt}), the second-order Lagrangian change in
vorticity is always different from zero for any values of the constants $A$
and $N$. Thus, at second-order, r-modes do not preserve vorticity of fluid
elements.

As it was mentioned in section \ref{lin}, the first-order canonical
displacements $\xi^{(1)i}$ are precisely those that preserve vorticity. As a
result, if the vorticity of a linear r-mode perturbation is initially small,
then as the perturbation grows under gravitational radiation reaction the
vorticity stays small. However, the situation changes when one takes into
account second-order effects; due to differential rotation an initially small
vorticity may increase as the perturbation grows.

\section{\label{ang}Angular momentum and energy of the r-mode perturbation}

The physical angular momentum of the second-order r-mode solution found in the
previous section is, at lowest order in $\Omega$, given by
\begin{eqnarray}
\delta^{(2)} J &=& \int \rho \delta^{(2)} v_{\phi} dV
\nonumber
\\
&=& \frac12 (l-1)(2l+1) \alpha^2 \Omega
R^{2-2l} \int_0^R \rho r^{2l+2} dr
\nonumber
\\
&& {} + 2\pi \alpha^2 \Omega A \int_0^R \rho r^{N+3} dr \int_0^\pi
\sin^{N+2}\theta d\theta, \label{physangm}
\end{eqnarray}
where $\delta^{(2)} v_{\phi}$ is given by eqs.~(\ref{difrot3}) and
(\ref{stra3}).

Since the physical angular momentum depends on the arbitrary constants $A$ and
$N$, one would like, at this point, to specify some physical conditions that
fix the values of these constants, i.e., that pick out a member of the family
of the second-order solution given by equations (\ref{difrot}), (\ref{partU}),
(\ref{stra}), and (\ref{stra4}). Since in the linearized theory vorticity is
conserved, it would be natural to impose the condition that the Lagrangian
change in vorticity at every order should be zero. However, as we have seen,
due to the presence of differential rotation, vorticity is not conserved
already at second order.

The second-order physical angular momentum can be decomposed in two pieces
\cite{fs}; one linear in the second-order Lagrangian change in velocity
$\Delta_{\xi}^{(2)} v_i$ and another one, called the canonical angular
momentum $J_c$, quadratic in the first-order Lagrangian displacement vector
$\xi^{(1)i}$:
\begin{equation}
\delta^{(2)}J=\frac{1}{\Omega}\int\rho v^i \Delta_{\xi}^{(2)}v_i dV +J_c,
\label{angm}
\end{equation}
where
\begin{equation}
J_c = - \int \rho \partial_{\phi} \xi^{(1)i}
\left( \partial_{t} \xi^{(1)}_i +v^k \nabla_k \xi^{(1)}_i \right) dV
\end{equation}
and the second-order Lagrangian change in velocity is given by
eq.~(\ref{lc2v}).

For the Lagrangian displacement given by eqs.~(\ref{ld})
and eqs.~(\ref{edc})--(\ref{c}), the canonical angular momentum
and the integral of the second-order Lagrangian change in velocity
are computed to be, respectively,
\begin{equation}
J_c = -\frac14 l(l+1) \alpha^2 \Omega R^{2-2l} \int_0^R \rho r^{2l+2} dr
\label{canangm}
\end{equation}
and
\begin{eqnarray}
\frac{1}{\Omega}\int\rho v^i \Delta^{(2)}_{\xi}v_i dV &=& \frac14 \left(
5l^2-l-2 \right) \alpha^2 \Omega R^{2-2l} \int_0^R \rho r^{2l+2} dr \nonumber
\\
&& {} + 2\pi \alpha^2 \Omega A \int_0^R \rho r^{N+3} dr \int_0^\pi
\sin^{N+2}\theta d\theta.
  \label{intgama}
\end{eqnarray}

As it can be seen from the above equations, the second-order physical change
in the angular momentum is not equal, in general, to the canonical angular
momentum. Since the second-order physical change in the energy (in the
inertial frame) $\delta^{(2)} E$ and the canonical energy $E_c$ are related by
the expression \cite{fs}
\begin{equation}
\delta^{(2)}E=\int\rho v^i \Delta_{\xi}^{(2)}v_i dV +E_c,
\label{enif}
\end{equation}
$\delta^{(2)} E$ and $E_c$ are also, in general, not equal.

In recent years, the nonlinear behaviour of r-modes was investigated within a
toy model of a spherical shell of rotating incompressible fluid \cite{lu}. It
was shown that in a spherical shell r-modes carry zero physical angular
momentum, $\delta^{(2)}J=0$ \footnote{In a spherical shell, the r-mode
solution $v^{\theta} \propto \alpha \Omega \sin^{l-1}\theta \sin(l\phi+\omega
t)$ and $v^{\phi} \propto \alpha \Omega \sin^{l-2}\theta \cos\theta
\cos(l\phi+\omega t)$, which is exact for arbitrary amplitude of the mode,
gives a zero contribution to the physical angular momentum of the shell. In a
full star, the above solution (with $v^r=0$) is just a first order
approximation and one has to take into account higher order terms, the
axisymmetric part of which does contribute to the physical angular momentum of
the star.}, and positive physical energy, $\delta^{(2)}E>0$, while both the
canonical angular momentum and canonical energy are negative. Thus, for
r-modes in a spherical shell one cannot equate physical and canonical
quantities. Based on these results, it was conjectured in Ref.~\cite{lu} that
also in the case of a full star physical angular momentum and energy of
r-modes are not equal to the corresponding canonical quantities. Our
investigation, on r-modes at second order in $\alpha$, confirms that for a
full star, \textit{in general}, $\delta^{(2)}J\neq J_c$ and $\delta^{(2)}E\neq
E_c$. Note, however, that for specific choices of initial data [such that the
two terms on the right hand side of eq.~(\ref{intgama}) cancel each other]
physical and canonical quantities can be made equal.

\section{\label{dis}Discussion and conclusions}

We have found a nonlinear extension of the linear r-mode perturbation,
describing differential rotation of pure kinematic nature that produces large
scale drifts along stellar latitudes. Differential rotation of fluid elements
given by eqs.~(\ref{difrot}) is induced by first-order quantities, while
differential rotation given by eqs.~(\ref{stra}) is a pure second-order
effect. The latter is stratified on cylinders and, therefore, cannot cancel
differential rotation induced by first-order quantities, which is not
stratified on cylinders. As already mentioned, our computation was carried out
in a slow rotation approximation, $\Omega \ll \Omega_K$; in order to neglect
higher-order terms arising from the expansion in powers of the star's angular
velocity $\Omega$, it was also assumed that $\alpha \gg (\Omega/\Omega_K)^2$
[see discussion before eq.~(\ref{eulerp2a})]. Recent results indicate that in
rapidly rotating neutron stars (just born in a supernova or spun up by
accretion in low-mass X-ray binaries) the r-mode instability may saturate at
low values \cite{rls,afmstw}, implying that $\alpha$ could be of the same
order of magnitude as $(\Omega/\Omega_K)^2$. In that case, terms in
$\delta^{(2)} v^i$ proportional to $\alpha\Omega^3$ (arising in an expansion
in powers of the angular velocity of the star) are as important as terms
proportional to $\alpha^2\Omega$ (arising in an expansion in powers of the
mode amplitude) and, therefore, should be taken into account. However, even
though the velocity field $\delta^{(2)}v^i$ and other second-order quantities
derived in this paper are, strictly speaking, only valid in the regime $\Omega
\ll \Omega_K$ and $\alpha \gg (\Omega/\Omega_K)^2$, they could be used to
illustrate the influence of differential rotation of r-modes on the nonlinear
evolution of rapidly rotating neutron stars.

Recently, an analytical expression for the azimuthal drift velocity of the
$l=2$ mode was derived from the linearized fluid equations by expanding the
velocity of a fluid element located at a certain point in powers of $\alpha$,
averaging over a gyration, and retaining only the lowest-order nonvanishing
term \cite{rls}. A comparison with our results shows the expression obtained
in Ref.~\cite{rls} is qualitatively correct but not exact to ${\cal
O}(\alpha^2)$. This is to be expected since the procedure used there does not
consider nonlinear effects in the fluid equations.

In the linearized theory, r-mode perturbation preserves vorticity of fluid
elements, with the consequence that the vorticity will not grow as the
perturbation grows under gravitational radiation reaction. However, the
situation changes when one takes into account second-order effects; due to
differential rotation of fluid elements, producing large scale azimuthal
drifts, the second-order Lagrangian change in vorticity is always different
from zero and may increase as the perturbation grows under gravitational
radiation reaction.

It was also explicitly shown that for r-modes physical and canonical
quantities cannot be equated. Canonical angular momentum (or energy) is not
the full angular momentum (or energy) at second order; one should also include
a part linear in the second-order Lagrangian change in velocity, which, as
pointed out in Ref.~\cite{fs}, is related to conservation of circulation in
the fluid. Since, as shown above, at second-order r-modes do not conserve
vorticity, it follows that, in general, the physical and canonical angular
momentum (or energy) do not coincide. However, specific choices of $A$ and $N$
can be made such that the integral in the right-hand side of eqs.~(\ref{angm})
and (\ref{enif}) vanishes. Such a case (for $l=2$) was studied in
Ref.~\cite{olcsva} within a phenomenological model for the evolution of
r-modes. However, it is not clear which physical condition would force the
constants $A$ and $N$ to take such particular values; thus, it seems more
appropriate to study the evolution of r-modes for the case of arbitrary values
of $A$ and $N$.

In recent years, numerical simulations have been used to study the nonlinear
evolution of r-modes, both for relativistic and Newtonian stars. Initial data
for this evolution was generated by adding a linear perturbation $\delta^{(1)}
v^i$ to an equilibrium stellar model. An ana\-ly\-tical expression for
$\delta^{(1)} v^i$ is known only for small-amplitude perturbations of
slowly-rotating Newtonian stars. As shown in this paper, the linear r-mode
perturbations given by eq.~(\ref{vf}) induce, at second order in the mode
amplitude, a drift of fluid elements along stellar latitudes, which cannot be
avoided by special choices of the constants $A$ and $N$. Thus, initial data
for numerical non-linear evolution of r-modes should also include a piece
describing differential rotation.

\begin{acknowledgments}
I would like to thank Nils Andersson, Curt Cutler, John Friedman, Keith
Lockitch, Benjamin Owen, Alberto Vecchio, and especially Bernard Schutz for
very helpful comments and discussions. The generous hospitality of the Albert
Einstein Institute in Golm, where part of this work was done, is acknowledged.
This work was supported in part by the \textit{Funda\c c\~ao para a Ci\^encia
e a Tecnologia}, Portugal.
\end{acknowledgments}

\bibliography{citac2}

\end{document}